\documentclass[12pt]{elsarticle}

\usepackage{subcaption}
\usepackage[utf8]{inputenc}
\usepackage[export]{adjustbox}
\usepackage{wrapfig}
\usepackage{graphicx}
\usepackage{amssymb}
\usepackage{amsmath, nccmath}
\usepackage{commath}
\usepackage{amsthm}
\usepackage{geometry}
\usepackage{lineno}

\usepackage{lipsum}
\makeatletter
\def\ps@pprintTitle{%
 \let\@oddhead\@empty
 \let\@evenhead\@empty
 \def\@oddfoot{}%
 \let\@evenfoot\@oddfoot}
\makeatother

\begin{document}

\begin{frontmatter}



\title{Casimir interactions from infinite range and dilation symmetry}



\author{Venkat Abhignan, R. Sankaranarayanan}

\address{Department of Physics, National Institute of Technology, Tiruchirappali - 620015, India.}

\begin{abstract}
The Casimir interaction energy for a class of discrete self-similar configuration of parallel plates is evaluated using existing methods. The similarities to characteristics of an attractive Casimir force is deduced only at infinite range of configuration. Further, the emergence of Casimir-like energy is qualitatively described for a Gaussian model of Landau-Ginzburg scalar field. Its relevance to self-similarity in the statistical field is shown at infinite range of fluctuations.
\end{abstract}


\end{frontmatter}

\section{Introduction}
Casimir interaction was initially studied as an attractive force between two perfectly conducting metal plates \cite{Casimir:1948dh}. Casimir interpreted that this interaction is dependent only on the distance between the plates, and electromagnetic fluctuations whose wavelength are comparable with the distance between the plates would be contributing to the Casimir effect. Casimir energy density or Casimir force per unit area is derived to be a negative quantity which is non-intuitive; however, since the Casimir energy produced outside the plates from an infinite range of fluctuations is more than inside, it is interpreted to be an attractive force between the plates. Typically, Casimir energy is derived with reference to quantum vacuum in field theories \cite{Miloni1994,Milton2001}. The range of fluctuations associated with zero-point energies in the vacuum is infinity. Apart from the electromagnetic field, scalar fields are also employed to obtain the Casimir effect. Any physical phenomenon associated with the dynamic interaction of fields with boundaries, non-trivial topology or space-time curvature can produce the Casimir effect \cite{Bordag:2009zz, Milton_2004,PLUNIEN1986,Carlos2006,RevModPhys.81.1827}. Casimir energy density associated to thermal Casimir effect \cite{theory0} are computed from polarization effects of vacuum in scalar fields of nontrivial space-time topology \cite{ford1975,FORD1976,DEWITT1975,vstft,Bezerra2011}. In this case, due to the absence of plates, boundary conditions are realized through identification conditions. Casimir interactions are also realized without discussing the zero-point energies \cite{scalar2,scalar3,scalar4,SCHWINGER1978,zpe1} and by using toy models \cite{zpe2,zpe3}, but infinite sums are used. Critical Casimir effect is studied with boundary conditions in the vicinity of second-order phase transitions using the theory of critical phenomenon \cite{fisher,theory1,RevModPhys1999,Krech,theory2}. Casimir-like effects arise in such statistical field theories when fluctuations are spatially confined close to the criticality. Such theories employ phenomenological Landau-Ginzburg parameters which are independent of microscopic properties dependent only on coarse-grained nature of the field with symmetries. The range of fluctuations in such critical systems is defined by correlation length, which tends to infinity. This work attempts to qualitatively relate the infinite range of self-similar fluctuations in a scalar field with the emergence of Casimir-like behaviour. 

In Sec.2, we briefly describe the nature of negative sums as seen when summing infinite range for Casimir energy \cite{Casimir:1948dh} and show its relation with self-similarity. Further, we use self-similar relations to derive Casimir energy for a self-similar configuration of plates using existing methods \cite{Shajesh2016}. In Sec. 3, we handle the Gaussian model of Landau-Ginzburg scalar field \cite{kardar2007statistical} to derive Casimir-like interaction. 
\section{Casimir interaction for self-similar configuration of parallel plates}
 A self-similar behaviour can be observed in an infinite series such as \begin{equation}
    \sum_{i=0}^{\infty}a_i x^i
\end{equation} where $\{a_i\}$ are constants, $x$ is considered to be an integer and $x>1$. One simple way to realise the sum of this divergent series ($S = \sum_{i=0}^{\infty} a_i x^i$) in case of $a_{i+1}/a_{i} \geq 1$ is by analytic continuation using self-similar function \cite{Yukalov1991,PhysRevE.55.6552,YUKALOV2002,physics3040053} such as continued fraction \cite{bender1999advanced,LORENTZEN20101364,abhignan2020continued} \begin{equation}
    \sum_{i=0}^{\infty}a_i x^i \sim \frac{b_0}{\frac{b_1 x}{\frac{b_2 x}{\cdots}+1}+1}
\end{equation} and observing its sequence of partial sums \begin{equation}
   \frac{b_0}{b_1 x + 1}, \frac{b_0}{\frac{b_1 x}{b_2 x+1}+1},\cdots
\end{equation} converge to a meaningful value called as a regularized sum. For a simple instance of $\{a_i\}=a$ in series (1) the partial sums in Eq. (3) approach the value \begin{equation}
    a \sum_{i=0}^{\infty} x^i \sim \frac{a}{1-x}.
\end{equation} For any value of $x>1$ this sum is negative, which is a non intuitive result in real plane. However, this sum is much more meaningful in complex $x$ plane, where its analogy can be seen in a Mercator projection as displayed in Fig.1.
\begin{figure}[h]
\caption{Mercator projection}
\centering
\includegraphics[width=0.4\textwidth]{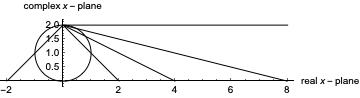}
\end{figure}
As shown, every point on the real plane can be projected on a unit circle of the complex plane by a line connecting the point and the top of the circle. As the values tend to infinity, the projection reaches the parallel line to the real plane, beyond which the projection is towards the negative real plane. This is just an analogy that shows the projection towards the negative values by finding a self-similar nature in the series, which is performed by the self-similar function. Another simple way to deduce the sum of this infinite series in Eq.(4) is to identify a self-similar relation if possible such as \begin{equation}
    S = a+x S 
\end{equation} and solve $S=a/(1-x)$. \\ Using self-similar relations, Casimir interaction energies were solved for a self-similar configuration of infinite sequence of parallel Dirac $\delta$-function plates \cite{Shajesh2016}. These infinitely thin plates with large masses (to neglect dynamics), interact with a quantum scalar field potential $V_i(\mathbf{x})=\lambda_i(z-a_i)$ placed at $z=a_1,a_2,a_3,\cdots$. Previously, Casimir energy was solved for configuration of plates placed at $z=2a,4a,6a,\cdots$ that is self-similar. Similarly, the result is extended to infinite sequence of self-similar plates placed at $z=x a,x^2 a,\cdots$ ($x>1$) using the idea of self-similarity. \\ For two parallel plates separated by distance $a$, the total energy per unit area $\varepsilon$, is decomposed to independent energies as \cite{casind1,casind2,casind3,casind4,Shajesh2016} \begin{equation}
    \varepsilon = \varepsilon_0 + \Delta\varepsilon_1+\Delta\varepsilon_2+\Delta\varepsilon(a).
\end{equation}
Here $\varepsilon_0$ is the energy of background vacuum in absence of plates, $\Delta\varepsilon_1$ and $\Delta\varepsilon_2$ are independent energies. $\Delta\varepsilon(a)$ is the Casimir interaction energy and depends only on the distance between the plates $a$. In general, the Casimir interaction energy per unit area between two plates satisfying Dirichlet boundary conditions is given by \cite{Milton2001} \begin{equation}
    \Delta\varepsilon_{12}(a)=-\frac{\pi^2}{1440a^3}.
\end{equation} in natural units which is used henceforth. This interaction energy for a configuration of plates placed at  $z=x a,x^2 a,\cdots$ can be solved by taking independent energies of two sub-systems at $z=x a$ and $z=x^2 a,x^3 a,\cdots$ as 
\begin{equation}
    \left( \varepsilon_0 + \sum_{i=1}^{\infty} \Delta\varepsilon_i+\Delta\varepsilon(x a) \right)=\varepsilon_0+\Delta\varepsilon_1+ \left( \sum_{i=2}^{\infty} \Delta\varepsilon_i+\Delta\varepsilon(x^2 a)+ \right)+\Delta\varepsilon_{12}(x^2 a -x a)
\end{equation}
Using the form of self-similarity, one can eliminate independent contributions of energies ($\{\Delta\varepsilon_i\}$) and obtain \begin{equation}
    \Delta\varepsilon(x a)=\Delta\varepsilon(x^2 a)+\Delta\varepsilon_{12}(x^2 a -x a)
\end{equation} Since distance between the plates $a$ is the only parameter in this system of configuration, based on dimensional grounds \begin{equation}
   \frac{1}{x^3} \Delta\varepsilon(x a) = \Delta\varepsilon(x^2 a) 
\end{equation}  and Eq. (9) can be solved as \begin{equation}
    \Delta\varepsilon(x a)= \frac{1}{x^3} \Delta\varepsilon(x a)-\frac{\pi^2}{1440(x^2 a-x a)^3}
\end{equation} to obtain Casimir interaction energy per unit area for complete configuration at $z=x a,x^2 a,\cdots$ as \begin{equation}
    \Delta\varepsilon(x a)= -\frac{\pi^2}{1440a^3(x-1)^3(x^3-1)}.
\end{equation} This result shows that for any value of $x>1$ at infinite range, this self-similar configuration of plates have an attractive force which corresponds with Casimir's result: where two conducting plates at distance $a$ attract each other under electromagnetic fluctuations with force per unit area or with energy density as \cite{Casimir:1948dh} \begin{equation}
    \left(\frac{E}{V}\right)_C= -\frac{\partial \varepsilon}{\partial a} = -\frac{\pi^2}{240 a^4}\,\, \hbox{for}\,\, \varepsilon = -\frac{\pi^2}{720 a^3} \hbox{(exactly twice of Eq.(7))}.
\end{equation} We now consider a self-similar configuration of plates placed at $z=a, a/x,a/x^2,\cdots$ in finite range. Similarly, using the idea of self-similarity one can obtain expression such as \begin{equation}
    \Delta\varepsilon(a)=\Delta\varepsilon(a/x)+\Delta\varepsilon_{12}(a -a/x)
\end{equation} by considering two sub-systems at $z=a$ and $z=a/x,a/x^2,\cdots$. And then solve for Casimir interaction energy per unit area for the entire configuration as \begin{equation}
    \Delta\varepsilon(a)= \frac{\pi^2 x^3}{1440a^3(x-1)^3(x^3-1)}.
\end{equation} In contrast, this result is always a repulsive force for all $x>1$. These results can also be checked by placing the two stacks of configurations $z=a, a/x,a/x^2,\cdots$ and $z=x a,x^2 a,\cdots$ next to each other, and solving for Casimir interaction energy per unit area for this sequence of plates which is exactly zero due to inflation nature of first stack and contraction nature of second stack \cite{Shajesh2016}. The total Casimir energy $\Delta\varepsilon_{tot}(a)$ of configuration $z=a, a/x,a/x^2,x a,x^2 a,\cdots$ can be solved from their individual Casimir components where \begin{equation}
   \Delta\varepsilon_{tot}(a)= \Delta\varepsilon(a) +\Delta\varepsilon(x a)+\Delta\varepsilon_{12}(x a - a). 
\end{equation} Using their independent expressions we have \begin{equation}
   \Delta\varepsilon_{tot}(a)=\frac{\pi^2 x^3}{1440a^3(x-1)^3(x^3-1)}-\frac{\pi^2}{1440a^3(x-1)^3(x^3-1)}-\frac{\pi^2}{1440a^3(x - 1)^3} = 0
\end{equation}
\section{Casimir-like effect from self-similar fluctuations in Landau-Ginzburg Gaussian theory}
The most general form of the Landau-Ginzburg Hamiltonian \cite{kardar2007statistical} with consideration of translational and rotational symmetries for a scalar field $\phi(\mathbf{x})$ is \begin{equation}
    \beta H = \int \hbox{d}^d\mathbf{x} \left[ \frac{t}{2} \phi^2 (\mathbf{x}) + u \phi^4 (\mathbf{x}) + v \phi^6 (\mathbf{x}) + \cdots+ \frac{K}{2}(\nabla \phi)^2 + \frac{L}{2}(\nabla^2 \phi)^2 + \cdots\right]. \label{23}
    \end{equation} This can cater for studying a wide range of physical systems undergoing second-order phase transitions at critical temperature $T=T_c$. The fluctuations in field $\phi(\mathbf{x})$ play an important role at the point of transition and they are governed by correlation functions such as $\left<| \phi(\mathbf{x})-\phi(\mathbf{x}')|^2\right>$ which is proportional to the characteristic length scale of the system $\xi$, the correlation length. At the point of criticality, $\xi \rightarrow \infty$ and the correlation functions $F_{crit}$ have a scale invariant or self-similar behaviour $ F_{crit}(\lambda \mathbf{x})= \lambda^e F_{crit}(\mathbf{x})$. The Hamiltonian or moreover the partition function of critical system expanded by a factor $\lambda$ possesses same statistical properties of the original one. This is signature of dilation symmetry in continuous field theories whereas fractal symmetry is more in relevance to discrete models. \\ We handle the Gaussian model of this Hamiltonian to realise a Casimir-like effect from a simple renormalization procedure similar to our previous work \cite{Abhignan2021}. This process is analogous to renormalization procedure proposed by Lifshitz for two thick plates separated by distance $a$ \cite{Lifshitz:1956zz, Bordag:2009zz}. The Casimir free energy $(E_c)$ of fluctuating electromagnetic field is obtained by subtracting free energy $(\Delta E_0(a))$ at limiting case where $a \rightarrow \infty$ $(\Delta E_{0,a \rightarrow \infty})$. Here $\xi$, the correlation length of the fluctuating scalar field is taken analogous to $a$. Gaussian model is solved by taking into account only the quadratic terms which gives the partition function as
     \begin{equation}
    Z = \int D\phi(\mathbf{x}) \exp\left(- \int \hbox{d}^d\mathbf{x} \left[ \frac{t}{2} \phi^2 (\mathbf{x}) + \frac{K}{2}(\nabla \phi)^2 + \frac{L}{2}(\nabla^2 \phi)^2 + \cdots\right]\right),
 \end{equation}
 where $\int D \phi(\mathbf{x})$ indicates integrating over all allowed configurations of the scalar field. Here $t$, $K$ and $L$ are Landau-Ginzburg parameters and analytical functions of temperature $T$ which can be expanded around critical temperature $T_c$ in Taylor series such as  \begin{align}
 t(T)=t_1 (T-T_c)+t_2 (T-T_c)^2+\cdots,
     \\ K(T)=K_0+K_1 (T-T_c)+K_2 (T-T_c)^2+ \cdots 
     \\ \hbox{and}\,\,\,\,L(T)=u_0+L_1 (T-T_c)+L_2 (T-T_c)^2+ \cdots .
     \end{align}  As observed in Eq.(20), $t$ is monotonic function of $T$ and at lowest orders, close to $T_c (t \rightarrow 0)$ \begin{equation}
         \frac{\partial Z}{\partial T} \sim \frac{\partial Z}{\partial t}\,\, \hbox{and energy}\,\,E \,\,\hbox{of this system is given as}\,\, E = T^2 \frac{\partial \ln Z}{\partial t}.
     \end{equation} 
 Considering Fourier modes of the fluctuating scalar field $$\phi(\mathbf{x})=\sum_{\mathbf{q}} \frac{1}{V}\phi_\mathbf{q} e^{i\mathbf{q}.\mathbf{x}},$$
 and re-expressing the partition function in terms of Fourier modes we have
  \begin{equation}
    Z \sim \int d\phi_\mathbf{q} \exp\left\{-\sum_{\mathbf{q}} \left(\frac{t+K q^2+L q^4 + \cdots}{2V}\right) |\phi_\mathbf{q}|^2 \right\}. \label{zgauss}
 \end{equation} The evaluation of different terms in above integral are as follows;
 \begin{multline}
     \int \hbox{d}^d\mathbf{x} (\nabla\phi)^2 = \int \hbox{d}^d\mathbf{x} \nabla\left( \sum_{\mathbf{q}} \frac{1}{V}\phi_\mathbf{q} e^{i\mathbf{q}.\mathbf{x}} \right) \nabla\left( \sum_{\mathbf{q}'} \frac{1}{V}\phi_\mathbf{q}' e^{i\mathbf{q}'.\mathbf{x}} \right) \\ 
     = \frac{1}{V^2}\sum_{\mathbf{q}} \sum_{\mathbf{q}'} \phi_\mathbf{q} \phi_{\mathbf{q}'}(-\mathbf{q}\mathbf{q}')\int \hbox{d}^d\mathbf{x}  \left( e^{i(\mathbf{q}+\mathbf{q}').\mathbf{x}} \right).
  \end{multline}
  \begin{equation} \hbox{ Since}\,\,
   \int \hbox{d}^d\mathbf{x} \left(e^{i(\mathbf{q}+\mathbf{q}').\mathbf{x}}\right) = V\delta_{\mathbf{q}+\mathbf{q}',0} \,\,\hbox{,}\,\, \int \hbox{d}^d\mathbf{x} (\nabla\phi)^2 = \frac{1}{V} \sum_{\mathbf{q}} q^2\abs{\phi_\mathbf{q}^2}.   \end{equation} 
   \begin{equation}
       \hbox{Similarly} \int \hbox{d}^d\mathbf{x} \phi^2 = \frac{1}{V} \sum_{\mathbf{q}} \abs{\phi_\mathbf{q}^2} .  
   \end{equation}
Primarily, the coarse-graining part of renormalization procedure involves subdividing the fluctuations into two components as \begin{equation}
\phi_\mathbf{q} = \Bigg\{\begin{array}{l}
   \ \widetilde{\phi_\mathbf{q}} \ \ \hbox{for}\ 0<\mathbf{q}< \Lambda/b \\ 
  \ \sigma_\mathbf{q} \ \  \hbox{for}\ \Lambda/b<\mathbf{q}< \Lambda
  \end{array}.
  \end{equation} Here the scalar field coarse-grained over $\Lambda/b$ from $\mathbf{q}=0$ consist of modes close to $\xi \rightarrow \infty$, which is responsible for the singularities. And we consider that large modes of fluctuations ($\Lambda/b<\mathbf{q}< \Lambda$) at short distances are responsible for Casimir-like interaction. We take $\Lambda/b$ corresponds to longest fluctuations influencing the plates while $\Lambda$ corresponds to shortest fluctuations affecting the plates for $b>1$. Considering the continuous modes of $\mathbf{q}$ we obtain the decoupled partition function in Eq.(24) as \begin{multline}
  Z = \int \hbox{d}\widetilde{\phi_\mathbf{q}} \hbox{d}\sigma_\mathbf{q} \exp\left[-\int_{0}^{\Lambda/b}\frac{\hbox{d}^d\mathbf{q}}{(2\pi)^d} |\widetilde{\phi_\mathbf{q}}|^2\left(\frac{t+K q^2+L q^4 + \cdots}{2}\right) \right] \\ \exp\left[-\int_{\Lambda/b}^{\Lambda}\frac{\hbox{d}^d\mathbf{q}}{(2\pi)^d} |\sigma_\mathbf{q}|^2\left(\frac{t+K q^2+L q^4 + \cdots}{2}\right) \right].
  \end{multline}
The analytic part of this integration, the orthogonal Gaussian modes of $\sigma_\mathbf{q}$ can be evaluated as \begin{multline}
  \ln Z= \ln\left(\exp\left[-\frac{V}{2}\int_{\Lambda/b}^{\Lambda} \frac{\hbox{d}^d\mathbf{q}}{(2\pi)^d} \ln(t+K q^2+L q^4 + \cdots)\right]\right) \\  + \ln \left( \int \hbox{d}\widetilde{\phi_\mathbf{q}} \exp\left[-\int_{0}^{\Lambda/b}\frac{\hbox{d}^d\mathbf{q}}{(2\pi)^d} |\widetilde{\phi_\mathbf{q}}|^2\left(\frac{t+K q^2+L q^4 + \cdots}{2}\right) \right]\right).
  \end{multline}
  The partition function for $\widetilde{\phi_\mathbf{q}}$ is self-similar to original Gaussian partition function in Eq. (24) when we consider rescaled modes $\mathbf{q'}=b\mathbf{q}$ and renormalized scalar field $\phi'(\mathbf{q'})$=
  $\phi(\mathbf{q'})/B$, which results in \begin{multline}
  \ln Z=-\frac{V}{2}\int_{\Lambda/b}^{\Lambda} \frac{\hbox{d}^d\mathbf{q}}{(2\pi)^d} \ln(t+K q^2+L q^4 + \cdots) \\ + \ln \left( \int d\widetilde{\phi'_\mathbf{q'}} \exp\left[-\int_{0}^{\Lambda}\frac{d^d\mathbf{q'}}{(2\pi)^d} b^{-d} B^2 |\widetilde{\phi'_\mathbf{q'}}|^2\left(\frac{t+K b^{-2} q'^2+L b^{-4} q'^4 + \cdots}{2}\right) \right]\right).
  \end{multline}
  The latter part of this partition function holds the infinite behaviour where exact self-similarity is for $b \gtrapprox 1$ and $B \approx 1$ mapping parameters $K,L,\cdots$ to themselves, while the initial finite part corresponding to Casimir-like energy using Eq.(23) is  \begin{equation}
      \left(\frac{E}{V}\right)_C = -\frac{ T^2}{2}\int_{\Lambda/b}^{\Lambda} \frac{\hbox{d}^d\mathbf{q}}{(2\pi)^d} \frac{1}{(t+K q^2+L q^4 + \cdots)}.
  \end{equation}
  In the Fourier space, this corresponds to a hypervolume of a thin shell between two hyperspheres with fluctuations ranging between $\Lambda/b$ to $\Lambda$. Similarly, the second term of partition function with the infinite range of fluctuations in Eq. (30) corresponds to hypersphere with radius $\Lambda/b$. Further, the leading dependence of fluctuations is evaluated on the energy of this system in Eq.(32) by considering $\mathbf{q}=\sqrt{t/K}x$ and $k_d=S_d/(2\pi)^d$, where $S_d$ is the solid angle in $d$ dimensions for spherically symmetric Fourier modes ($\hbox{d}^d\mathbf{q}=S_d\mathbf{q}^{(d-1)}$). \begin{equation}
    \left(\frac{E}{V}\right)_C = -\frac{k_d}{2}\left(\frac{t}{K}\right)^{d/2}\frac{T^2}{t}\int_{\frac{\Lambda\sqrt{K}}{b\sqrt{t}}}^{\Lambda\sqrt{K/t}}  \frac{x^{(d-1)}\hbox{d}x}{(1+x^2+Lt x^4/K^2 + \cdots)}.
    \end{equation}
    This expression shows the leading behaviour of fluctuations for arbitrary $d$ as \begin{equation}
    \left(\frac{E}{V}\right)_C\propto -\frac{\Lambda^d}{b^d}[b^d-1] .\end{equation} For $b \gtrapprox 1$ which can relate to thin plates and $b/\Lambda$ corresponding to the distance between the plates $(b/\Lambda \sim a)$, this expression gives the standard Casimir interaction behaviour $(E/V)_c \propto -a^{-d}$ \cite{Milton_2004}. For $d=4$ the behaviour is similar to physically relevant case in Eq.(13). This can imply that self-similar behaviour of the scalar field fluctuations in this Gaussian model at an infinite range leads to Casimir-like behaviour. While we studied this critical system close to $T=T_c$, further the role of temperature can be studied by systematically involving $t,L,K,\cdots$. 

\section{Conclusion}
     Casimir interaction energy is derived for a class of self-similar configuration of $\delta$-plates with quantum scalar field potential and its relation to attractive Casimir force is related to infinite range of configuration. Gaussian model of Landau-Ginzburg scalar field is considered and renormalization procedure was implemented to show emergence of Casimir-like energy under self-similar behaviour of the statistical field with infinite range of fluctuations. These results may highlight the relationship between infinite range of fluctuations, self-similarity and physically relevant attractive Casimir force. We have only qualitatively found these relations for approximate models. Further, taking a general realistic system and finding their exact relationship for Casimir interaction would be interesting. The relation between self-similar nature of electromagnetic field and an attractive Casimir force was previously proposed and quantitatively studied in a revised model of electromagnetism \cite{funaro2009fractal}. This was explored using Maxwell’s equations embedded in the framework of general relativity \cite{funaro1,funaro2}.




\bibliographystyle{model1-num-names}
\bibliography{sample.bib}

\begin{thebibliography}{44}
\expandafter\ifx\csname natexlab\endcsname\relax\def\natexlab#1{#1}\fi
\providecommand{\bibinfo}[2]{#2}
\ifx\xfnm\relax \def\xfnm[#1]{\unskip,\space#1}\fi
\bibitem[{Casimir(1948)}]{Casimir:1948dh}
\bibinfo{author}{H.~B.~G. Casimir},
\newblock \bibinfo{title}{{On the Attraction Between Two Perfectly Conducting
  Plates}},
\newblock \bibinfo{journal}{Indag. Math.} \bibinfo{volume}{10}
  (\bibinfo{year}{1948}) \bibinfo{pages}{261--263}. \bibinfo{note}{[Kon. Ned.
  Akad. Wetensch. Proc.100N3-4,61(1997)]}.
\bibitem[{Milonni(1994)}]{Miloni1994}
\bibinfo{editor}{P.~W. Milonni} (Ed.), \bibinfo{title}{The Quantum Vacuum},
  \bibinfo{publisher}{Academic Press}, \bibinfo{address}{San Diego},
  \bibinfo{year}{1994}.
\bibitem[{Milton(2001)}]{Milton2001}
\bibinfo{author}{K.~A. Milton}, \bibinfo{title}{The Casimir Effect},
  \bibinfo{publisher}{WORLD SCIENTIFIC}, \bibinfo{year}{2001}.
\bibitem[{Bordag et~al.(2009)Bordag, Klimchitskaya, Mohideen, and
  Mostepanenko}]{Bordag:2009zz}
\bibinfo{author}{M.~Bordag}, \bibinfo{author}{G.~L. Klimchitskaya},
  \bibinfo{author}{U.~Mohideen}, \bibinfo{author}{V.~M. Mostepanenko},
  \bibinfo{title}{{Advances in the Casimir effect}}, volume
  \bibinfo{volume}{145}, \bibinfo{publisher}{Oxford University Press},
  \bibinfo{year}{2009}.
\bibitem[{Milton(2004)}]{Milton_2004}
\bibinfo{author}{K.~A. Milton},
\newblock \bibinfo{title}{The casimir effect: recent controversies and
  progress},
\newblock \bibinfo{journal}{Journal of Physics A: Mathematical and General}
  \bibinfo{volume}{37} (\bibinfo{year}{2004}) \bibinfo{pages}{R209--R277}.
\bibitem[{Plunien et~al.(1986)Plunien, Müller, and Greiner}]{PLUNIEN1986}
\bibinfo{author}{G.~Plunien}, \bibinfo{author}{B.~Müller},
  \bibinfo{author}{W.~Greiner},
\newblock \bibinfo{title}{The casimir effect},
\newblock \bibinfo{journal}{Physics Reports} \bibinfo{volume}{134}
  (\bibinfo{year}{1986}) \bibinfo{pages}{87--193}.
\bibitem[{Farina(2006)}]{Carlos2006}
\bibinfo{author}{C.~Farina},
\newblock \bibinfo{title}{The casimir effect: some aspects},
\newblock \bibinfo{journal}{Brazilian Journal of Physics} \bibinfo{volume}{36}
  (\bibinfo{year}{2006}) \bibinfo{pages}{1137–1149}.
\bibitem[{Klimchitskaya et~al.(2009)Klimchitskaya, Mohideen, and
  Mostepanenko}]{RevModPhys.81.1827}
\bibinfo{author}{G.~L. Klimchitskaya}, \bibinfo{author}{U.~Mohideen},
  \bibinfo{author}{V.~M. Mostepanenko},
\newblock \bibinfo{title}{The casimir force between real materials: Experiment
  and theory},
\newblock \bibinfo{journal}{Rev. Mod. Phys.} \bibinfo{volume}{81}
  (\bibinfo{year}{2009}) \bibinfo{pages}{1827--1885}.
\bibitem[{Brown and Maclay(1969)}]{theory0}
\bibinfo{author}{L.~S. Brown}, \bibinfo{author}{G.~J. Maclay},
\newblock \bibinfo{title}{Vacuum stress between conducting plates: An image
  solution},
\newblock \bibinfo{journal}{Phys. Rev.} \bibinfo{volume}{184}
  (\bibinfo{year}{1969}) \bibinfo{pages}{1272--1279}.
\bibitem[{Ford(1975)}]{ford1975}
\bibinfo{author}{L.~H. Ford},
\newblock \bibinfo{title}{Quantum vacuum energy in general relativity},
\newblock \bibinfo{journal}{Phys. Rev. D} \bibinfo{volume}{11}
  (\bibinfo{year}{1975}) \bibinfo{pages}{3370--3377}.
\bibitem[{Ford(1976)}]{FORD1976}
\bibinfo{author}{L.~H. Ford},
\newblock \bibinfo{title}{Quantum vacuum energy in a closed universe},
\newblock \bibinfo{journal}{Phys. Rev. D} \bibinfo{volume}{14}
  (\bibinfo{year}{1976}) \bibinfo{pages}{3304--3313}.
\bibitem[{DeWitt(1975)}]{DEWITT1975}
\bibinfo{author}{B.~S. DeWitt},
\newblock \bibinfo{title}{Quantum field theory in curved spacetime},
\newblock \bibinfo{journal}{Physics Reports} \bibinfo{volume}{19}
  (\bibinfo{year}{1975}) \bibinfo{pages}{295 -- 357}.
\bibitem[{Dowker and Critchley(1977)}]{vstft}
\bibinfo{author}{J.~S. Dowker}, \bibinfo{author}{R.~Critchley},
\newblock \bibinfo{title}{Vacuum stress tensor in an einstein universe:
  Finite-temperature effects},
\newblock \bibinfo{journal}{Phys. Rev. D} \bibinfo{volume}{15}
  (\bibinfo{year}{1977}) \bibinfo{pages}{1484--1493}.
\bibitem[{Bezerra et~al.(2011)Bezerra, Klimchitskaya, Mostepanenko, and
  Romero}]{Bezerra2011}
\bibinfo{author}{V.~B. Bezerra}, \bibinfo{author}{G.~L. Klimchitskaya},
  \bibinfo{author}{V.~M. Mostepanenko}, \bibinfo{author}{C.~Romero},
\newblock \bibinfo{title}{Thermal {Casimir} effect in closed {Friedmann}
  universe revisited},
\newblock \bibinfo{journal}{Phys. Rev. D} \bibinfo{volume}{83}
  (\bibinfo{year}{2011}) \bibinfo{pages}{104042}.
\bibitem[{Schwinger(1975)}]{scalar2}
\bibinfo{author}{J.~Schwinger},
\newblock \bibinfo{title}{Casimir effect in source theory},
\newblock \bibinfo{journal}{Letters in Mathematical Physics}
  \bibinfo{volume}{1} (\bibinfo{year}{1975}) \bibinfo{pages}{43--47}.
\bibitem[{Schwinger(1992{\natexlab{a}})}]{scalar3}
\bibinfo{author}{J.~Schwinger},
\newblock \bibinfo{title}{Casimir effect in source theory ii},
\newblock \bibinfo{journal}{Letters in Mathematical Physics}
  \bibinfo{volume}{24} (\bibinfo{year}{1992}{\natexlab{a}})
  \bibinfo{pages}{59--61}.
\bibitem[{Schwinger(1992{\natexlab{b}})}]{scalar4}
\bibinfo{author}{J.~Schwinger},
\newblock \bibinfo{title}{Casimir effect in source theory iii},
\newblock \bibinfo{journal}{Letters in Mathematical Physics}
  \bibinfo{volume}{24} (\bibinfo{year}{1992}{\natexlab{b}})
  \bibinfo{pages}{227--230}.
\bibitem[{Schwinger et~al.(1978)Schwinger, DeRaad, and Milton}]{SCHWINGER1978}
\bibinfo{author}{J.~Schwinger}, \bibinfo{author}{L.~L. DeRaad},
  \bibinfo{author}{K.~A. Milton},
\newblock \bibinfo{title}{Casimir effect in dielectrics},
\newblock \bibinfo{journal}{Annals of Physics} \bibinfo{volume}{115}
  (\bibinfo{year}{1978}) \bibinfo{pages}{1 -- 23}.
\bibitem[{Jaffe(2005)}]{zpe1}
\bibinfo{author}{R.~L. Jaffe},
\newblock \bibinfo{title}{Casimir effect and the quantum vacuum},
\newblock \bibinfo{journal}{Phys. Rev. D} \bibinfo{volume}{72}
  (\bibinfo{year}{2005}) \bibinfo{pages}{021301}.
\bibitem[{Nikolić(2017)}]{zpe2}
\bibinfo{author}{H.~Nikolić},
\newblock \bibinfo{title}{Is zero-point energy physical? a toy model for
  {Casimir}-like effect},
\newblock \bibinfo{journal}{Annals of Physics} \bibinfo{volume}{383}
  (\bibinfo{year}{2017}) \bibinfo{pages}{181 -- 195}.
\bibitem[{Nikolić(2016)}]{zpe3}
\bibinfo{author}{H.~Nikolić},
\newblock \bibinfo{title}{Proof that {Casimir} force does not originate from
  vacuum energy},
\newblock \bibinfo{journal}{Physics Letters B} \bibinfo{volume}{761}
  (\bibinfo{year}{2016}) \bibinfo{pages}{197 -- 202}.
\bibitem[{Fisher and de~Gennes(1978)}]{fisher}
\bibinfo{author}{M.~E. Fisher}, \bibinfo{author}{P.~G. de~Gennes},
\newblock \bibinfo{title}{Phenomena at the walls in a critical binary mixture},
\newblock \bibinfo{journal}{C. R. Acad. Sci. Paris} \bibinfo{volume}{B 287}
  (\bibinfo{year}{1978}) \bibinfo{pages}{207--209}.
\bibitem[{Krech and Dietrich(1992)}]{theory1}
\bibinfo{author}{M.~Krech}, \bibinfo{author}{S.~Dietrich},
\newblock \bibinfo{title}{Specific heat of critical films, the casimir force,
  and wetting films near critical end points},
\newblock \bibinfo{journal}{Phys. Rev. A} \bibinfo{volume}{46}
  (\bibinfo{year}{1992}) \bibinfo{pages}{1922--1941}.
\bibitem[{Kardar and Golestanian(1999)}]{RevModPhys1999}
\bibinfo{author}{M.~Kardar}, \bibinfo{author}{R.~Golestanian},
\newblock \bibinfo{title}{The ``friction'' of vacuum, and other
  fluctuation-induced forces},
\newblock \bibinfo{journal}{Rev. Mod. Phys.} \bibinfo{volume}{71}
  (\bibinfo{year}{1999}) \bibinfo{pages}{1233--1245}.
\bibitem[{Krech(1994)}]{Krech}
\bibinfo{author}{M.~Krech}, \bibinfo{title}{The Casimir Effect in Critical
  Systems}, \bibinfo{publisher}{WORLD SCIENTIFIC}, \bibinfo{year}{1994}.
\bibitem[{Gross et~al.(2017)Gross, Gambassi, and Dietrich}]{theory2}
\bibinfo{author}{M.~Gross}, \bibinfo{author}{A.~Gambassi},
  \bibinfo{author}{S.~Dietrich},
\newblock \bibinfo{title}{Statistical field theory with constraints:
  Application to critical {Casimir} forces in the canonical ensemble},
\newblock \bibinfo{journal}{Phys. Rev. E} \bibinfo{volume}{96}
  (\bibinfo{year}{2017}) \bibinfo{pages}{022135}.
\bibitem[{Shajesh et~al.(2016)Shajesh, Brevik, Cavero-Pel\'aez, and
  Parashar}]{Shajesh2016}
\bibinfo{author}{K.~V. Shajesh}, \bibinfo{author}{I.~Brevik},
  \bibinfo{author}{I.~Cavero-Pel\'aez}, \bibinfo{author}{P.~Parashar},
\newblock \bibinfo{title}{Casimir energies of self-similar plate
  configurations},
\newblock \bibinfo{journal}{Phys. Rev. D} \bibinfo{volume}{94}
  (\bibinfo{year}{2016}) \bibinfo{pages}{065003}.
\bibitem[{Kardar(2007)}]{kardar2007statistical}
\bibinfo{author}{M.~Kardar}, \bibinfo{title}{Statistical Physics of Fields},
  \bibinfo{publisher}{Cambridge University Press}, \bibinfo{year}{2007}.
\bibitem[{Yukalov(1991)}]{Yukalov1991}
\bibinfo{author}{V.~I. Yukalov},
\newblock \bibinfo{title}{Method of self‐similar approximations},
\newblock \bibinfo{journal}{Journal of Mathematical Physics}
  \bibinfo{volume}{32} (\bibinfo{year}{1991}) \bibinfo{pages}{1235--1239}.
\bibitem[{Yukalov and Gluzman(1997)}]{PhysRevE.55.6552}
\bibinfo{author}{V.~I. Yukalov}, \bibinfo{author}{S.~Gluzman},
\newblock \bibinfo{title}{Self-similar bootstrap of divergent series},
\newblock \bibinfo{journal}{Phys. Rev. E} \bibinfo{volume}{55}
  (\bibinfo{year}{1997}) \bibinfo{pages}{6552--6565}.
\bibitem[{Yukalov and Yukalova(2002)}]{YUKALOV2002}
\bibinfo{author}{V.~Yukalov}, \bibinfo{author}{E.~Yukalova},
\newblock \bibinfo{title}{Self-similar structures and fractal transforms in
  approximation theory},
\newblock \bibinfo{journal}{Chaos, Solitons and Fractals} \bibinfo{volume}{14}
  (\bibinfo{year}{2002}) \bibinfo{pages}{839 -- 861}. \bibinfo{note}{Fractal
  Geometry in Quantum Physics}.
\bibitem[{Yukalov and Yukalova(2021)}]{physics3040053}
\bibinfo{author}{V.~I. Yukalov}, \bibinfo{author}{E.~P. Yukalova},
\newblock \bibinfo{title}{From asymptotic series to self-similar approximants},
\newblock \bibinfo{journal}{Physics} \bibinfo{volume}{3} (\bibinfo{year}{2021})
  \bibinfo{pages}{829--878}.
\bibitem[{Bender and Orszag(1999)}]{bender1999advanced}
\bibinfo{author}{C.~Bender}, \bibinfo{author}{S.~Orszag},
  \bibinfo{title}{Advanced Mathematical Methods for Scientists and Engineers I:
  Asymptotic Methods and Perturbation Theory}, Advanced Mathematical Methods
  for Scientists and Engineers, \bibinfo{publisher}{Springer},
  \bibinfo{year}{1999}.
\bibitem[{Lorentzen(2010)}]{LORENTZEN20101364}
\bibinfo{author}{L.~Lorentzen},
\newblock \bibinfo{title}{Padé approximation and continued fractions},
\newblock \bibinfo{journal}{Applied Numerical Mathematics} \bibinfo{volume}{60}
  (\bibinfo{year}{2010}) \bibinfo{pages}{1364 -- 1370}.
  \bibinfo{note}{Approximation and extrapolation of convergent and divergent
  sequences and series (CIRM, Luminy - France, 2009)}.
\bibitem[{Abhignan and Sankaranarayanan(2021)}]{abhignan2020continued}
\bibinfo{author}{V.~Abhignan}, \bibinfo{author}{R.~Sankaranarayanan},
\newblock \bibinfo{title}{Continued functions and perturbation series: Simple
  tools for convergence of diverging series in {$O(n)$}-symmetric $\phi^4$
  field theory at weak coupling limit},
\newblock \bibinfo{journal}{Journal of Statistical Physics}
  \bibinfo{volume}{183} (\bibinfo{year}{2021}) \bibinfo{pages}{4}.
\bibitem[{Kenneth and Klich(2006)}]{casind1}
\bibinfo{author}{O.~Kenneth}, \bibinfo{author}{I.~Klich},
\newblock \bibinfo{title}{Opposites attract: A theorem about the casimir
  force},
\newblock \bibinfo{journal}{Phys. Rev. Lett.} \bibinfo{volume}{97}
  (\bibinfo{year}{2006}) \bibinfo{pages}{160401}.
\bibitem[{Shajesh and Schaden(2011)}]{casind2}
\bibinfo{author}{K.~V. Shajesh}, \bibinfo{author}{M.~Schaden},
\newblock \bibinfo{title}{Many-body contributions to green's functions and
  casimir energies},
\newblock \bibinfo{journal}{Phys. Rev. D} \bibinfo{volume}{83}
  (\bibinfo{year}{2011}) \bibinfo{pages}{125032}.
\bibitem[{Bulgac and Wirzba(2001)}]{casind3}
\bibinfo{author}{A.~Bulgac}, \bibinfo{author}{A.~Wirzba},
\newblock \bibinfo{title}{Casimir interaction among objects immersed in a
  fermionic environment},
\newblock \bibinfo{journal}{Phys. Rev. Lett.} \bibinfo{volume}{87}
  (\bibinfo{year}{2001}) \bibinfo{pages}{120404}.
\bibitem[{Bulgac et~al.(2006)Bulgac, Magierski, and Wirzba}]{casind4}
\bibinfo{author}{A.~Bulgac}, \bibinfo{author}{P.~Magierski},
  \bibinfo{author}{A.~Wirzba},
\newblock \bibinfo{title}{Scalar casimir effect between dirichlet spheres or a
  plate and a sphere},
\newblock \bibinfo{journal}{Phys. Rev. D} \bibinfo{volume}{73}
  (\bibinfo{year}{2006}) \bibinfo{pages}{025007}.
\bibitem[{Abhignan and Sankaranarayanan(2021)}]{Abhignan2021}
\bibinfo{author}{V.~Abhignan}, \bibinfo{author}{R.~Sankaranarayanan},
\newblock \bibinfo{title}{Casimir-like effect from thermal field fluctuations},
\newblock \bibinfo{journal}{Brazilian Journal of Physics} \bibinfo{volume}{51}
  (\bibinfo{year}{2021}) \bibinfo{pages}{1897--1903}.
\bibitem[{Lifshitz(1956)}]{Lifshitz:1956zz}
\bibinfo{author}{E.~M. Lifshitz},
\newblock \bibinfo{title}{{The theory of molecular attractive forces between
  solids}},
\newblock \bibinfo{journal}{Sov. Phys. JETP} \bibinfo{volume}{2}
  (\bibinfo{year}{1956}) \bibinfo{pages}{73--83}.
\bibitem[{Funaro(2009)}]{funaro2009fractal}
\bibinfo{author}{D.~Funaro},
\newblock \bibinfo{title}{The fractal structure of matter and the casimir
  effect},
\newblock \bibinfo{journal}{arXiv 0906.1874}  (\bibinfo{year}{2009}).
\bibitem[{Funaro(2008)}]{funaro1}
\bibinfo{author}{D.~Funaro}, \bibinfo{title}{Electromagnetism and the Structure
  of Matter}, \bibinfo{publisher}{WORLD SCIENTIFIC}, \bibinfo{year}{2008}.
\bibitem[{Funaro(2019)}]{funaro2}
\bibinfo{author}{D.~Funaro}, \bibinfo{title}{From Photons to Atoms},
  \bibinfo{publisher}{WORLD SCIENTIFIC}, \bibinfo{year}{2019}.

\end{thebibliography}







\end{document}